\pdfoutput=1
\documentclass{aa}  
\usepackage{graphicx,natbib}
\usepackage{caption}
\usepackage{subcaption}

\usepackage{txfonts}
\def\kms{km~s$^{-1}$}

\newcommand{\nc}{\newcommand}

\nc{\RAJ}[4]{$\alpha(J2000) = {#1}^{\rm h}{#2}^{\rm m}{#3}\fs{#4}$}
\nc{\DecJ}[4]{$\delta(J2000) = {#1}\degr {#2}\arcmin {#3}\farcs{#4}$}

\begin{document}

   \title{Polarisation of molecular lines in the circumstellar envelope of the post-Asymptotic Giant Branch star OH~17.7-2.0}

  \titlerunning{CO polarisation around OH~17.7-2.0}


   \author{W.~H.~T. Vlemmings
          \inst{1}\fnmsep\thanks{wouter.vlemmings@chalmers.se}
          \and
          D. Tafoya\inst{1}
          }

   \institute{Department of Space, Earth and Environment, Chalmers University of Technology, Onsala Space Observatory, 439 92 Onsala, Sweden
}

   \date{accepted 31-Jan-2023}

 
  \abstract
   {The role of magnetic field in the shaping of Planetary Nebulae (PNe), either directly or indirectly after being enhanced by binary interaction, has long been a topic of debate. Large scale magnetic fields around pre-PNe have been inferred from polarisation observations of masers. However, because masers probe very specific regions, it is still unclear if the maser results are representative of the intrinsic magnetic field in the circumstellar envelope (CSE).}
   {Molecular line polarisation of non-maser lines can provide important information about the magnetic field. A comparison between the magnetic field morphology determined from maser observations and that observed in the more diffuse CO gas, can reveal if the two tracers probe the same magnetic field.}
   {We compare observations taken with the Atacama Large Millimeter/submillimeter Array (ALMA) of molecular line polarisation around the post-Asymptotic Giant Branch (post-AGB)/pre-PNe star OH~17.7-2.0 with previous observations of polarisation in the 1612~MHz OH maser region. Earlier mid-infrared observations indicate that OH~17.7-2.0 is a young bipolar pre-PNe, with both a torus and bipolar outflow cavities embedded in a remnant AGB envelope.}
   {We detect CO~$J=2-1$ molecular line polarisation at a level of $\sim4\%$ that displays an ordered linear polarisation structure. We find that, correcting for Faraday rotation of the OH~maser linear polarisation vectors, the OH and CO linearly polarised emission trace the same large scale magnetic field. A structure function analysis of the CO linear polarisation reveals a plane-of-the-sky magnetic field strength of $B_\perp\sim1$~mG in the CO region, consistent with previous OH Zeeman observations.}
   {The consistency of the ALMA CO molecular line polarisation observation with maser observations indicate that both can be used to determine the magnetic field strength and morphology in CSEs. The new observations indicate that the magnetic field has a strong toroidal field component projected on the torus structure 
   and a poloidal field component along the outflow cavity. The existence of a strong, ordered, magnetic-field around OH 17.7-2.0 indicates that magnetic fields are likely involved in the formation of this bipolar pre-PNe.}


   \maketitle
%

\section{Introduction}
The processes involved in the formation of bipolar Planetary Nebulae (PNe) are still under debate \citep[e.g.][]{Blackman22}. While there are strong indications that binary interaction is required \citep[e.g.][and references therin]{Boffin19}, there are also observations that show that magnetic fields appear responsible for launching collimated outflows that shape the nebula \citep[e.g.][]{Vlemmings06}. It is possible that binarity and magnetic fields work in tandem. In this case, the magnetic field is enhanced by the interaction between the evolved star, typically during the red-giant branch (RGB) or asymptotic-giant branch (AGB) phase of their evolution, and a companion \citep[e.g.][]{Nordhaus06}. This interaction could lead to a common-envelope evolution (CEE), where the tenuous envelope of the RGB/AGB star is ejected while accretion on the companion or evolved star core, assisted by enhanced magnetic fields, launches a bipolar outflow \citep[e.g.][]{O22}. Observations of a large sample of evolved stars with fast bipolar outflows indicate that these bipolar pre-PNe likely make up the majority of the recent CEE ejection events \citep{Khouri22}.

\begin{figure*}
     \centering
     \begin{subfigure}[b]{0.49\textwidth}
         \centering
         \includegraphics[width=\textwidth]{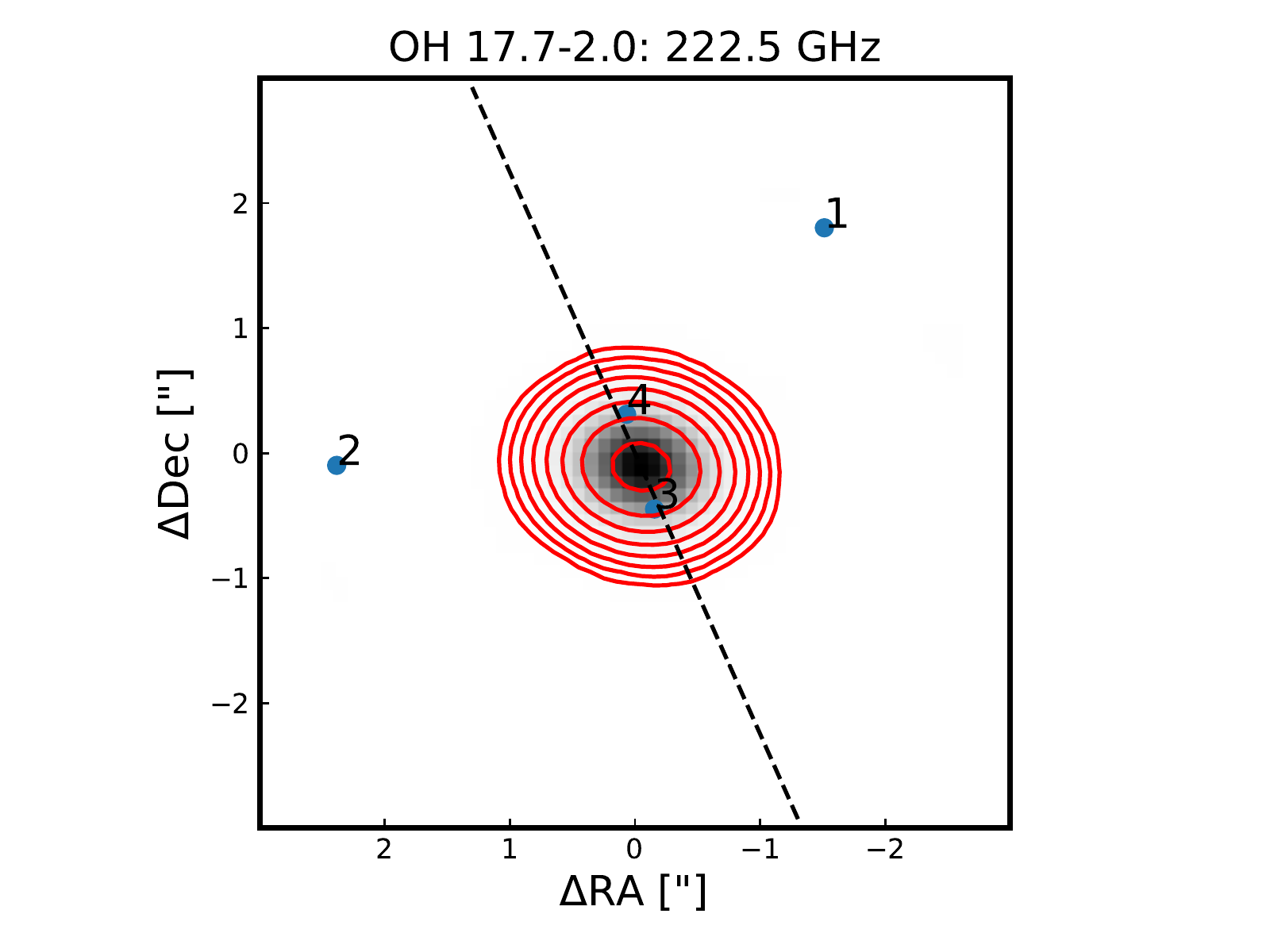}
     \end{subfigure}
     \hfill
     \begin{subfigure}[b]{0.49\textwidth}
         \centering
         \includegraphics[width=\textwidth]{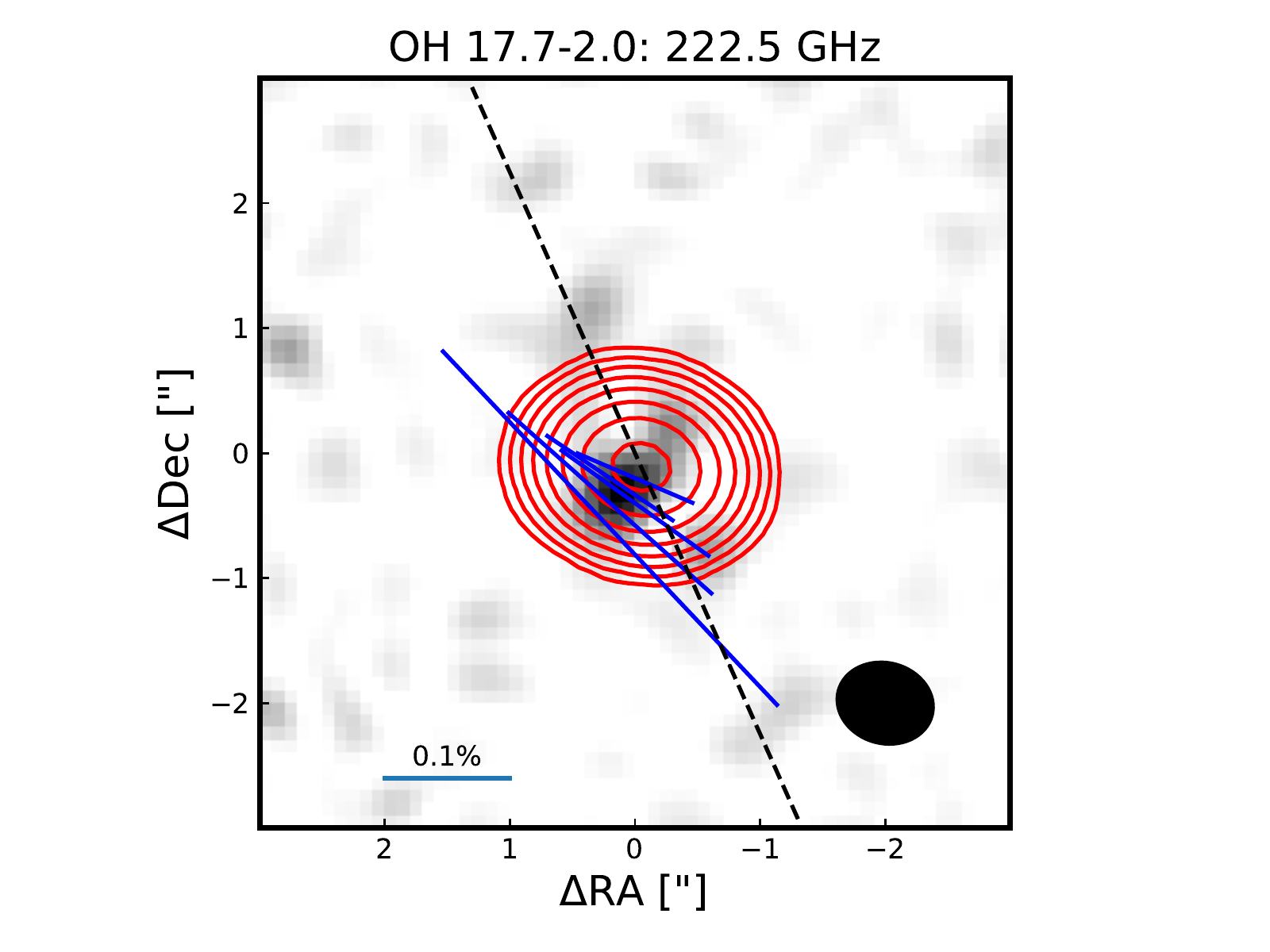}
     \end{subfigure}
     \hfill
        \caption{{\it (left)} The 222.5~GHz continuum emission from OH~17.7-2.0 in red contours and greyscale. The contours are drawn at $0.625, 1.25, 2.5, 5, 10, 20, 40,$ and $80\%$ of the peak value of 71.8~mJy~beam$^{-1}$. The dashed line indicates the direction of the outflow inferred from mid-infrared continuum and H$_2$ observations \citep{Gledhill11}. The four labelled solid circles are the sources identified by Gaia within a radius of $2.5$\arcsec of OH~17.7-2.0 (see text). {\it (right)} The polarised continuum emission (greyscale) and the 225.5~GHz continuum contours (similarly spaced as in the left panel) of OH~17.7-2.0. The blue line segments indicate the linear polarisation direction where polarised emission is detected at $>3\sigma$, where the rms on the polarised emission $\sigma=21~\mu$Jy~beam$^{-1}$. This means that the uncertainty on the direction is $\lesssim 10^\circ$. The segments are scaled by the linear polarisation fraction which peaks at $P_{\rm l,max}=0.47\%$. The dashed line indicates the direction of the outflow from OH~17.7-2.0 as described for the left panel. The filled ellipse indicates the beam size of the observations.}
        \label{fig:cont}
\end{figure*}

Strong magnetic fields have also been observed around early AGB stars and post-AGB/pre-PNe objects that have no confirmed companion or that have not yet undergone a CEE event \citep[e.g.][and references therein]{Vlemmings19}. The origin of these fields is still unclear.
The majority of the magnetic field observations around evolved stars have relied on measurements of linear and circular polarisation, due to the Zeeman-effect, of masers \citep[e.g.][]{Vlemmings02, Bains03, Vlemmings05, Vlemmings06, Herpin06, LealFerreira13, Gonidakis14}. Additionally, there are observations of the surface magnetic field on one AGB star \citep{Lebre14} as well as on the surface of two post-AGB stars \citep{Sabin15}. Observations of polarised dust \citep[e.g.][]{Sabin20} and radio synchrotron emission \citep{PerezSanchez13} around post-AGB stars might also indicate the involvement of magnetic fields in shaping PNe. Still, to date, the strongest evidence of the role of magnetic fields specifically around post-AGB objects comes from maser observations \citep[e.g.][]{Bains03, Vlemmings06} that provide both a magnetic field morphology and magnetic field strength. It has however been argued that the magnetic fields determined from maser observations are biased due to the special nature of the amplified stimulated emission that produces maser features in very specific region in the CSE \citep[e.g.][]{Soker02}. 

Molecular line polarisation from non-maser molecules, resulting from the Goldreich-Kylafis (GK) effect \citep[e.g.][]{GK82, Lankhaar20}, can be used to determine if the maser magnetic field observations probe the intrinsic large-scale magnetic field. 
The GK-effect of CO and other molecules has been observed in star forming regions, proto-planetary disks, and around a supergiant star \citep[e.g.][]{Cortes05, Beuther10, Vlemmings17, Stephens20, Cortes21, Teague21}. It has also been detected, using the Sub-Millimeter Array (SMA), around the AGB stars CW~Leo and IK Tau \citep{Girart12, Vlemmings12} and with the Combined Array for Research in Millimeter-wave Astronomy (CARMA) towards R~Leo and R~Crt \citep{Huang20}. While these observations indicated a possible large scale magnetic field, the sensitivity of the SMA and CARMA was not sufficient for the detection of polarised molecular line emission in more than a few compact regions of the CSE. For most of the previously observed sources, no maser polarisation observations probing the same region as probed by the GK-effect exist. It has thus not been possible to compare the observations directly with maser measurements.  

One of the post-AGB/pre-PNe sources with extensive polarised OH maser observations is the star OH~17.7-2.0. Polarisation observations of mainly the 1612~MHz OH masers around this source revealed an apparent coherent large scale structure with a magnetic field strength in the OH maser region of $B\approx2.5-4.5$~mG \citep[][herafter B03]{Bains03}. Here we present ALMA observations of linear polarisation due to the GK-effect in the envelope of OH~17.7-2.0. In \S~\ref{obs}, we present the observations and data reduction, and in \S~\ref{OH177} we present the source, with a particular discussion regarding its distance. In \S~\ref{res}, we present the results of the ALMA observations and in \S~\ref{disc} we perform a Structure Function Analysis (SFA) to estimate the magnetic field strength, perform a detailed comparison with the OH maser observation of B03, and present our results on the morphology of the magnetic field around OH~17.7-2.0. We end with our conclusions in \S~\ref{conc}.

\begin{table}
\caption{Gaia results}             
\label{Table:Gaia}      
\centering          
\begin{tabular}{c c c c }     
\hline\hline
 label & Gaia source\_id & $\pi\pm\sigma_\pi$ & offset \\
 & & [mas] & [\arcsec] \\
 \hline
1 & 4104128498748054272 & $0.39\pm0.02$ & $2.35$\\
2 & 4104128503103791232 & $-0.04\pm0.14$& $2.39$\\
3 & 4104128503103804288 & - & $0.47$\\
4 & 4104128503164186496 & - & $0.32$\\
\hline\hline       
\end{tabular}
\end{table}

\section{Observations and data reduction}
\label{obs}
The CSE of OH~17.7-2.0 was observed by ALMA in full polarisation mode on April 14 2018, spending a total time of 2.5~hours. The on-source time was $\sim70$~min. The remaining time was spent observing the phase calibrator J1832+2039 and the amplitude and polarisation calibrator J1924+2914. Four spectral windows of $1.875$~GHz and 960 channels each were centred on $214.6$, $216.5$, $228.6$, and $230.5$~GHz resulting in a channel width of $\sim2.6$~\kms. The observations were calibrated using the ALMA polarisation calibration scripts \citep{Nagai16}.

\begin{figure*}[ht!]
\centering
\includegraphics[width=0.95\textwidth]{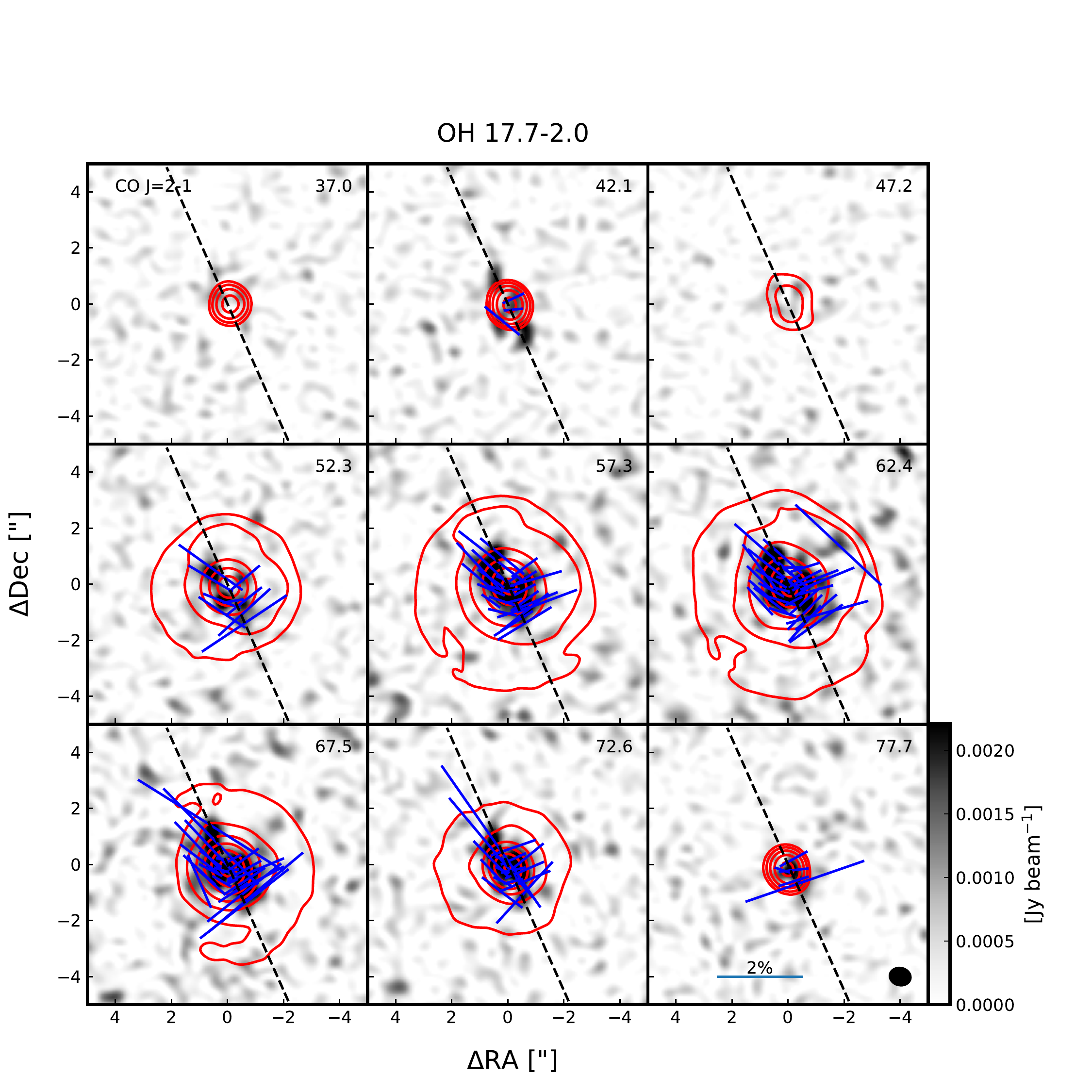}
\caption{Channel maps of the polarised CO $J=2-1$ emission around the post-AGB star OH~17.7-2.0. The Stokes I total intensity emission is indicated by the solid red contours at $2.5, 5, 10, 20, 40,$ and $80\%$ of the peak emission ($I_{\rm CO, peak}$=1.04~Jy~beam$^{-1}$). The linearly polarised emission is shown as a greyscale map, and the blue line segments denote the linear polarisation direction when the polarised emission $>5\sigma_P$. The uncertainty on the polarisation direction is thus $\lesssim 7^\circ$. The segments are scaled to the level of fractional polarisation with the scale indicated in the bottom left panel. The maximum polarisation $P_{\rm l, max}= 4.1\%$. The panels are labelled with the $V_{\rm lsr}$ velocity in \kms, and the beam size is shown in the bottom right panel. The stellar velocity $V_{\rm lsr, *}=62.0$~\kms. The dashed line indicates the direction of the outflow of OH~17.7-2.0 as in Fig.~\ref{fig:cont}}
\label{fig:CO}
\end{figure*}

 Subsequent calibration, self-calibration and imaging was done using CASA 5.7.2 \citep{McMullin07}\footnote{The data were checked against an error in visibility amplitude calibration (\url{https://almascience.eso.org/news/amplitude-calibration-issue-affecting-some-alma-data}) but ALMA staff from the ESO ALMA Regional Centre determined that a correction was not necessary}. We performed two rounds of phase-only self-calibration on the compact continuum of OH~17.7-2.0, which improved the dynamic range in the continuum from $\sim530$ to $\sim1010$. The observations reach a continuum rms in Stokes I ($\sigma_I$) of $71~\mu$Jy~beam$^{-1}$. In the Stokes Q and U continuum images, the rms noise ($\sigma_{Q,U}$) are $\sim 13~\mu$Jy~beam$^{-1}$ and $\sim 18~\mu$Jy~beam$^{-1}$ respectively. We also produced the Stokes V continuum image, for which we find an rms noise $\sigma_V\sim14~mu$Jy~beam$^{-1}$. While the rms in the Stokes Q, U and V images are close to the theoretical noise limit, that of Stokes I is almost five time higher. The continuum Stokes I image is thus likely limited by dynamic range. The continuum beam size, using Briggs weighing and a robust parameter of $0.5$, is $0.78\times0.65$\arcsec (PA $75.1^\circ$). 

The continuum was subtracted using the CASA task {\it uvcontsub} after which the strongest spectral lines were imaged using Briggs weighing and a robust parameter of $0.5$. To improve polarisation sensitivity, we averaged two channels, obtaining a velocity resolution of $\sim 5.2$~\kms. The $\sigma_I$, $\sigma_Q$, $\sigma_U$, and $\sigma_V$ rms noise level in a line free channel are $0.64~$mJy~beam$^{-1}$, $0.39~$mJy~beam$^{-1}$, $0.38~$mJy~beam$^{-1}$, and $0.47~$mJy~beam$^{-1}$. The beam size, at 230~GHz is $0.77\times0.64$\arcsec (PA $77.3^\circ$). The maximum recoverable scale in our observations is $\sim 8.1$\arcsec. From a comparison with the published CO $J=2-1$ spectrum \citep{Heske90} and a visual inspection of the images, there are no indications of significant resolved out flux. Finally linear polarisation maps were created from the Stokes Q and U imaged using $P_l=\sqrt{Q^2+U^2-\sigma^2_P}$. The polarisation rms $\sigma_P\approx0.5$~mJy~beam$^{-1}$ is found from an analysis of the rms in a line free spectral channel for each spectral line individually.

\section{OH~17.7-2.0}
\label{OH177}

 The post-AGB/pre-PNe star OH~17.7-2.0, also known as IRAS 18276-1431, has been studied across a wide range of wavelengths. It hosts strong OH masers \citep[e.g.][]{Bowers78}, while its H$_2$O masers disappeared after a rapid flux decline between 1985 and 1990 \citep{Engels02} before reappearing more than 20 years later \citep{Wolak13}. The decrease of H$_2$O maser emission was attributed to a recent drop in mass-loss rate caused by the star having recently left the AGB phase. Polarimetric measurements of OH masers appear to indicates that the CSE of OH~17.7-2.0 has a strong large-scale magnetic field (B03).
 OH~17.7-2.0 was first suggested be a bipolar nebula based on an analysis of its infrared (IR) spectral energy distribution \citep{LeBertre84}. Later imaging in the optical and (near-)IR confirmed this view \citep[e.g.][]{SC07, Gledhill11, Lagadec11, Murakawa13}. Observations of H$_2$ and CO emission in the near-IR suggest a bipolar outflow velocity of $\sim95$~\kms. The outflow has created bipolar outflow cavities that are $\gtrsim125$~yr old and are inclined to the line-of-sight by $\sim22^\circ$ \citep{Gledhill11}. Observations of the $^{12}$CO and $^{13}$CO $J=1-0$ emission using the Owens Valley Radio Observatory (OVRO) revealed that the source is embedded in a dense, mostly spherical, CO envelope which has an outflow velocity of $\sim17$~\kms \citep{SC07, SCS12}. No fast outflow is detected in the sub-millimetre CO emission. In \citet{Murakawa13}, a model was created to match the near-IR spectral energy distribution and polarised emission. They conclude, assuming a distance of $3$~kpc, that most of the dust resides in a torus with inner and outer radii of 30 and 1000~au. The torus has a total mass of $3.0$~M$_\odot$ and, assuming an expansion of the torus with a velocity similar to the velocities measured in the CO $J=1-0$, an age of $\sim300$~yr.   

{\bf The distance to OH~17.7-2.0:} Because foreground Faraday rotation of the linearly polarised OH maser emission affects the comparison of the absolute polarisation angle between the OH and sub-millimetre molecular line emission, we carefully assess the distance to OH~17.7-2.0. In the literature, distances are quoted between $\sim2-5.4$~kpc. Some of the earliest distances were derived using the phase-lag method, comparing the size of the OH maser shell with the delay between the infrared and OH maser variability curve. This yielded distances ranging from $3.4-5.4$~kpc using measurements from \citet{Bowers83} or $2-5.5$~kpc, using those from \citet{HH85}. Alternatively, the (near) kinematic distance, for a source velocity $V_{\rm LSR, *}=62$~\kms, gives a distance of $4.1^{+0.2}_{-0.3}$~kpc \citep{Reid14}. Finally, the luminosity distance, assuming a source luminosity of $6000$~L$_\odot$, is $2.94\pm0.38$~kpc \citep{Vickers15}. In the Gaia DR3 data release \citep{Gaia21}, four sources are identified within $2.5$\arcsec of the position of the sub-millimetre continuum peak found in our ALMA observations. The Gaia source identification number, offsets, and Gaia parallax results are shown in Table~\ref{Table:Gaia} and the positions are noted in Fig.~\ref{fig:cont}. It is clear that the two first Gaia sources are not related to OH~17.7-2.0. In fact, their positions coincide exactly with the field sources seen in the near-infrared observations of \citet{SC07} and \citet{Gledhill11}. The two remaining Gaia sources, correspond exactly with the scattered light emission coming from the two outflow lobes of OH~17.7-2.0 seen in the same observations. However, unsurprisingly, no reliable astrometry could be performed on these two lobes. Based on these results, we adopt the luminosity distance from \citet{Vickers15} of $D=2.94\pm0.38$~kpc as the most reliable at the moment.        

\section{Observational results}
\label{res}

We detected significant ($>3\sigma$) polarised emission towards the 222.5~GHz continuum of OH~17.7-2.0. The maximum fraction of continuum polarisation is $0.47\%$. This is a lower fraction than the continuum polarisation detected with ALMA towards the post-AGB object OH~231.8+4.2 \citep{Sabin20} (at $\sim345$~GHz). We detect no circular polarisation signal in the Stokes V continuum image, so can place a $3\sigma$ limit on the continuum circular polarisation of $0.58\%$.

Molecular line linear polarisation is detected (at $>5\sigma$) for four of the five strongest molecular transitions detected in our observations. The five transitions, all detected with a peak emission $>0.1$~Jy~beam$^{-1}$ are presented with their peak fluxes and maximum linear polarisation fraction (and $5\sigma$ polarisation upper limit, towards the peak of the line emission, in case of the non-detection) in Table~\ref{Table:res}. Channel maps of the four lines for which polarisation was detected, CO~$J=2-1$, $^{29}$SiO~$J=5-4$, SiO~$J=5-4$, and SO~$J=5-4$ are shown in Figs.~\ref{fig:CO}, \ref{fig:29SiO}, ~\ref{fig:SiO}, and~\ref{fig:SO54} respectively. In addition to the linear polarisation, we also investigated if circular polarisation we detected for any of the molecular lines. Compact, negative, Stokes V emission was detected in four channels towards the peak of the Stokes I emission of CO~$J=2-1$. The (absolute) maximum fractional circular polarisation $P_{\rm v, max}=0.41\%$. The other lines did not display significant Stokes V emission. Currently, the minimum detectable degree of circular polarisation, defined as three times the systematic calibration uncertainty, is $1.8\%$\footnote{ALMA Cycle 9 Technical handbook}, compared to $0.1\%$ for linear polarisation. As the measured level of the CO circular polarisation is below this, the circular polarisation is likely a result of the systematic calibration uncertainties. Still, we shortly discuss the measurement in \S~\ref{ARS} in relation to possible contributions from anisotropic resonance scattering \citep{Houde13, Houde22}.

\begin{table}
\caption{OH~17.7-2.0 line polarisation}             
\label{Table:res}      
\centering          
\begin{tabular}{l c  c }     
\hline\hline
Molecular Line & $I_{\rm peak}$ & $P_{\rm l, max}$ \\
 & [Jy~beam$^{-1}$] & [$\%$] \\
 \hline
$^{12}$CO $J=2-1$ & 1.04 & 4.1 \\
$^{29}$SiO $v=0, J=5-4$ & 0.31 & 4.7 \\
SiO $v=0, J=5-4$ & 0.48 & 2.9 \\
p-H$_2$S $(2_{2,0}-2_{1.1})$ & 0.29 & <0.5 \\
SO $J=5-4$ & 0.12 & 2.2 \\
\hline\hline       
\end{tabular}
\end{table}

\begin{figure*}[ht!]
\centering
\includegraphics[width=0.85\textwidth]{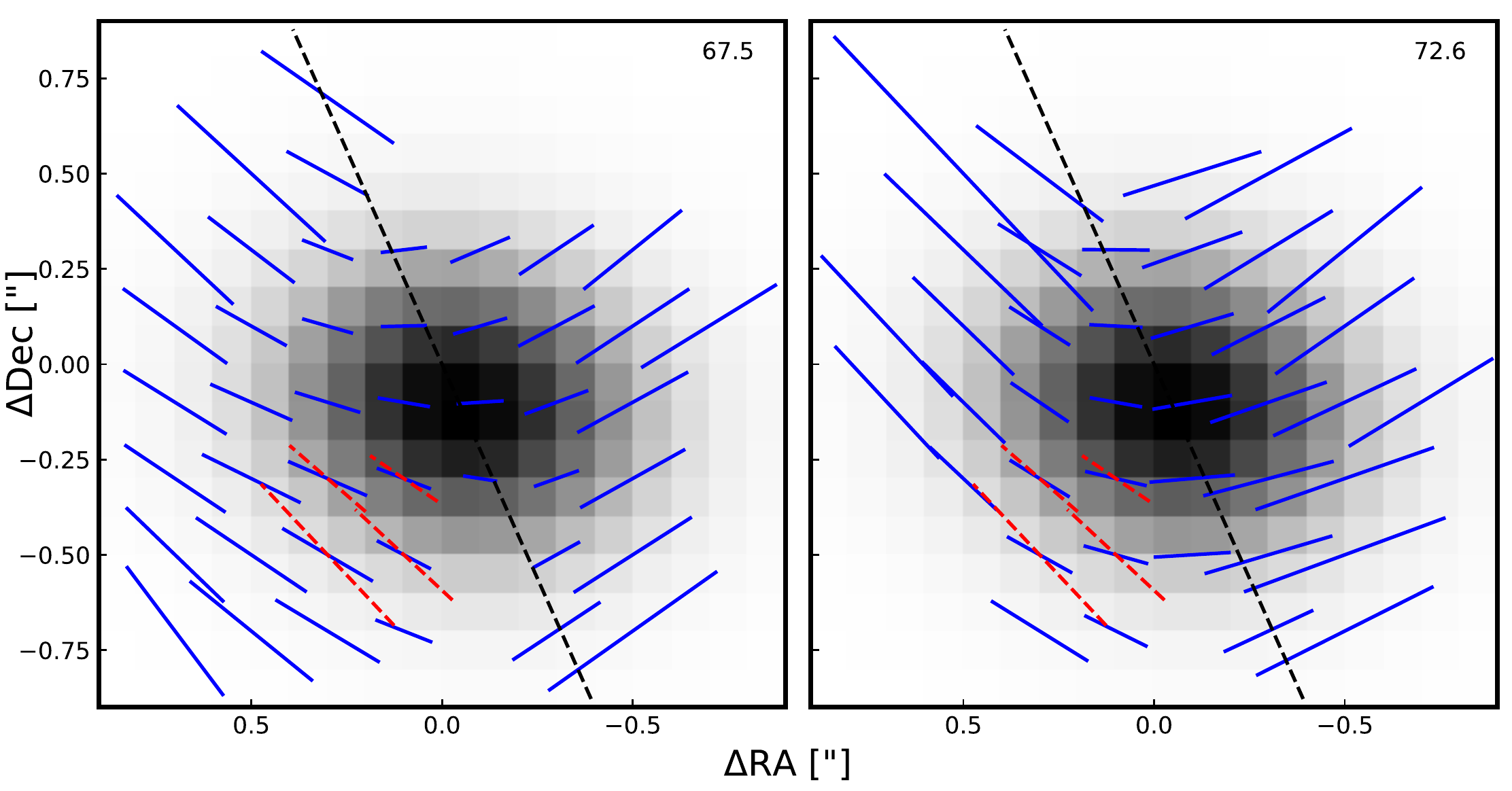}
\caption{A zoom-in on the continuum emission (grey scale) of OH~17.7-2.0 with the blue line segments denoting the linear polarisation of CO~$J=2-1$ for the two channels around the systemic velocity (labelled in km~s$^{-1}$ in the top right corner). The dashed red line segments indicate the linear polarisation direction of the continuum polarisation. The black dashed line indicates the direction of the bipolar outflow.}
\label{fig:dustcomp}
\end{figure*}

\begin{figure*}[ht!]
\centering
\includegraphics[width=0.95\textwidth]{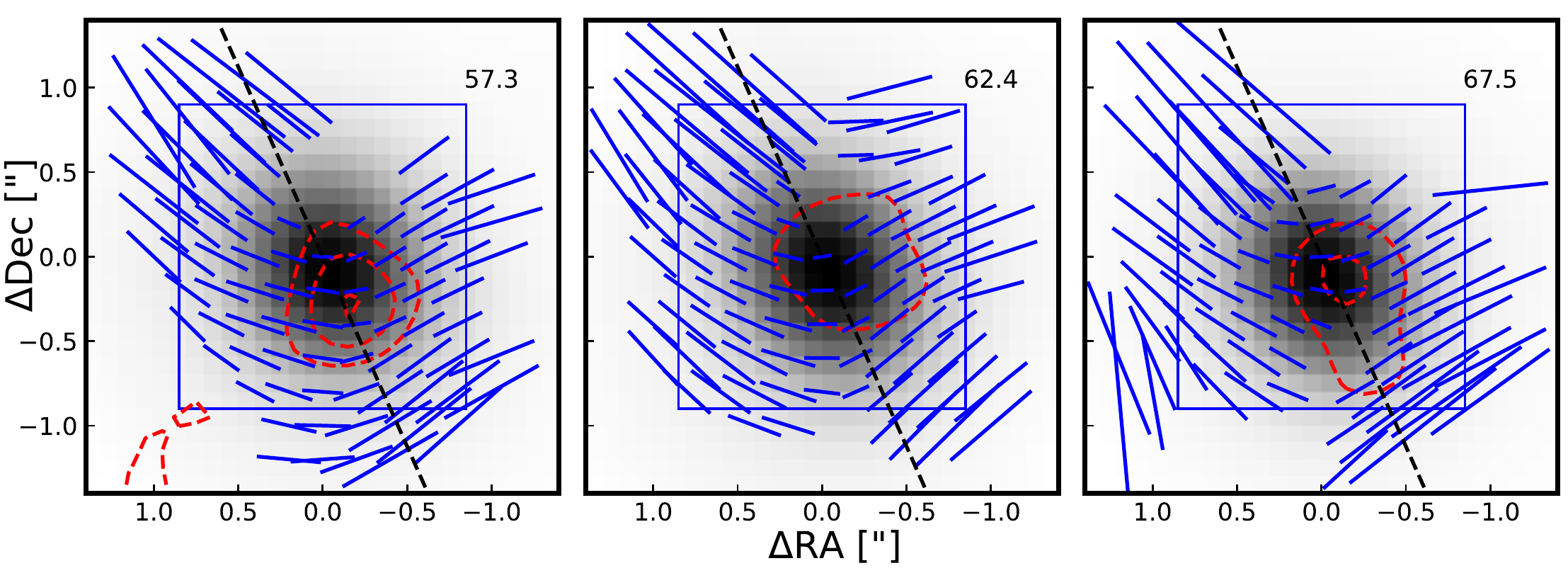}
\caption{A zoom-in to the central region of the CO~$J=2-1$ emission previously shown in Fig.~\ref{fig:CO}. The greyscale indicates the Stokes I emission and the blue line segments indicate the linearly polarisation direction. The red dashed contours represent the (negative) Stokes V emission at $-8, -6, -4\sigma$ where $\sigma=0.4$~mJy~beam$^{-1}$. The outflow direction is indicated by the dashed line. The blue box indicates the region of the 1612~MHz OH masers for which a linearly polarisation map is presented in B03 (their figure 7). The majority of the OH masers is located within $\sim0.4$\arcsec of the centre of the map. The channel velocity in in km~s$^{-1}$ is given the top right corner of each panel.} 
\label{fig:zoom}
\end{figure*}

\begin{figure*}
     \centering
     \begin{subfigure}[b]{0.49\textwidth}
         \centering
         \includegraphics[width=\textwidth]{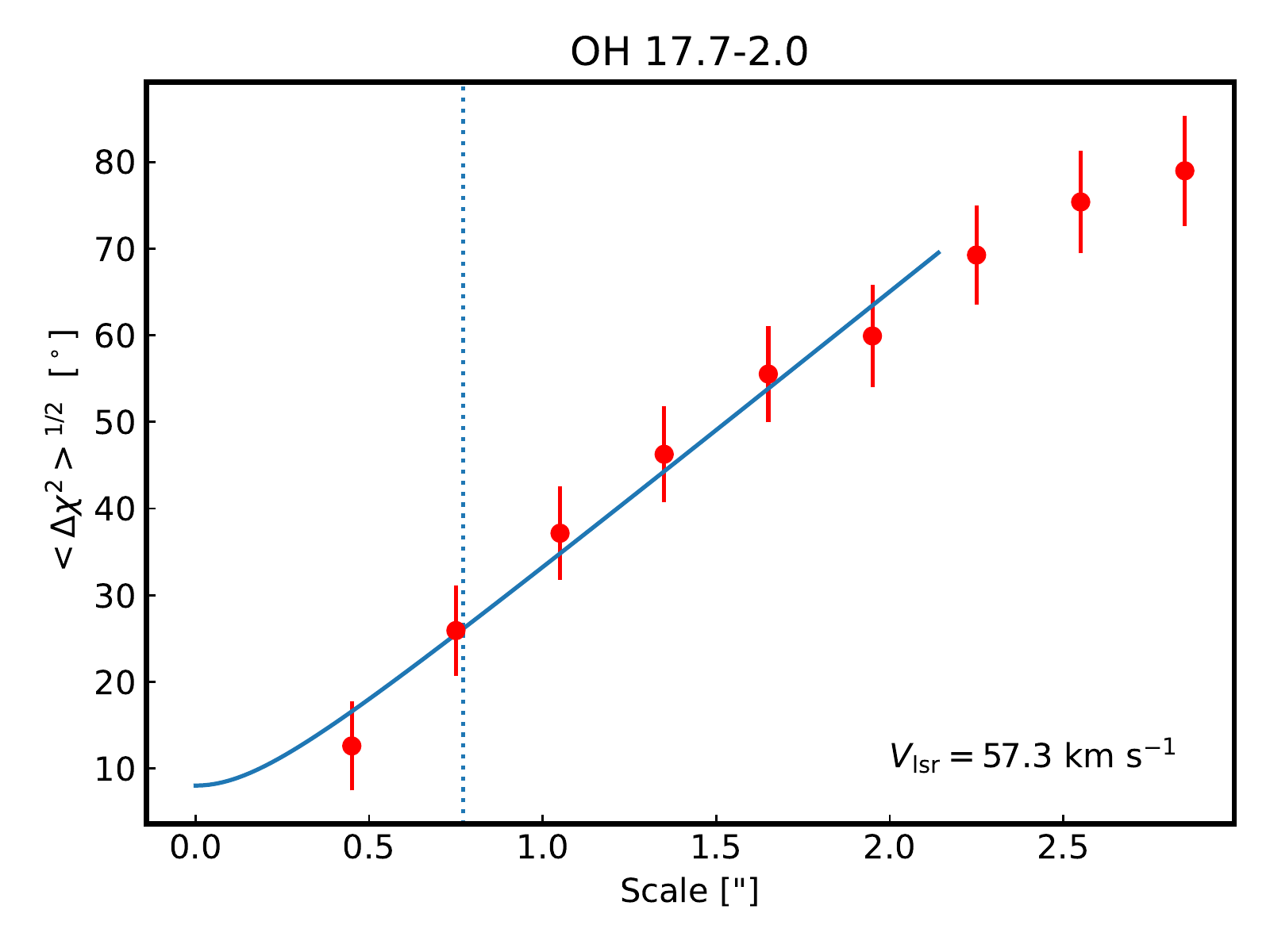}
     \end{subfigure}
     \hfill
     \begin{subfigure}[b]{0.49\textwidth}
         \centering
         \includegraphics[width=\textwidth]{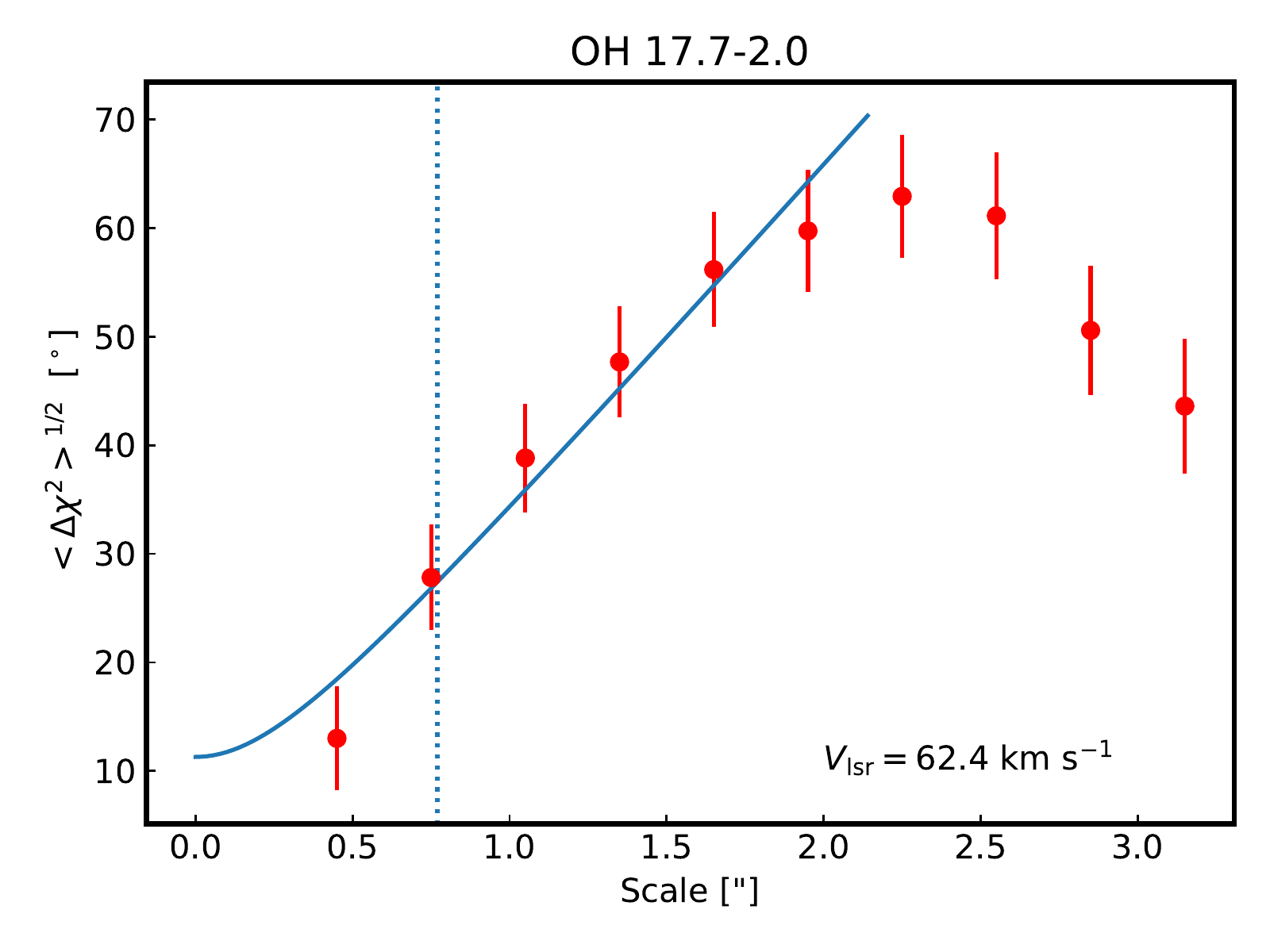}
     \end{subfigure}
     \hfill
        \caption{The dispersion of the polarisation vectors (the square root of the second order structure function), binned to the Nyquist sampled resolution, for two velocity channels of the CO $J=2-1$ polarised emission around OH~17.7-2.0. The error bars indicate the variance in each bin. The vertical dotted line indicates the size of the beam major axis. The solid line indicates the fit of the Structure Function Analysis described in the text.}
        \label{fig:SFAOH}
\end{figure*}

Only for the CO $J=2-1$ do we detect significant polarisation across several consecutive velocity channels and in extended regions of more than $3$\arcsec in size. The polarisation direction for the CO appears to curve from an angle of $\sim 120^\circ$ towards the South along the outflow axis, to $\sim 35^\circ$ towards the North. This means it goes from being nearly perpendicular to the outflow to being nearly parallel to the outflow. In contrast, the $^{29}$SiO $v=0, J=5-4$ has, at blue-shifted velocities, polarisation vectors that lie at and angle of approximately $45^\circ$ with respect to the outflow axis. Around $V_{\rm lsr}=68.5$~\kms, the detected polarisation is perpendicular to the outflow. Polarisation of SiO $v=0, J=5-4$ is only detected in one area, at $V_{\rm lsr}=68.3$~\kms, towards the North on the outflow axis, with the polarisation perpendicular to the outflow. Finally, the SO $J=5-4$ also only displays one region of significant polarisation, at $V_{\rm lsr}=56.1$~\kms, with a direction similar to that of the CO at the same velocity.

\section{Discussion}
\label{disc}

\subsection{Continuum polarisation}
\label{continuum}

As seen in Fig.~\ref{fig:cont}(right) the continuum polarisation is confined to a small area slightly offset from the continuum peak. While the central star contributes to the emission, the largest contribution comes from the circumstellar dust. Considering our spatial resolution is not sufficient to resolve details in the continuum emission, and there are no observations at other frequencies that can be used to constrain the dust properties or the alignment mechanism, we do not provide an in depth analysis of the dust polarisation. As shown in e.g.\citet{Vlemmings17}, for the supergiant VY~CMa, the dust polarisation is likely due to alignment of the dust particles with the magnetic field. In Fig.~\ref{fig:dustcomp}, we compare the continuum polarisation direction with those of the CO~$J=2-1$ emission in spectral channels close to the systemic velocity. As seen in the comparison, the direction of the polarisation is completely consistent within the polarisation direction uncertainties, indicating that the origin of the polarisation in the molecular lines and continuum is related.

\subsection{Anisotropic resonant scattering}
\label{ARS}
 Anisotropic resonant scattering can affect the polarisation properties of molecular line emission when it passes through a magnetised molecular region in the foreground between the source and the telescope \citep{Houde13, Houde22}. It can also occur in the emitting molecular line region itself. This process results in a transformation from linear to circular polarisation. Consequently, the linear polarisation direction is potentially no longer directly relatable to the magnetic field direction. 

 Since the emitting CO region around evolved stars is limited by CO photo-dissociation in the outer CSE, there is unlikely to be a large column of CO molecules in the foreground towards the source. Additionally, the large velocity gradient in an expanding CSE, towards all but the extreme velocities, further reduces the foreground column of gas at the emitting velocity towards most lines of sight. Thus, anisotropic resonance scattering of the polarised CO line emission from CSEs is likely limited to the scattering that occurs in the emitting region itself.  

 If anisotropic resonance scattering affects the linearly polarised emission detected in our observations, we expect a component of circular polarisation in the affected regions. Although, as described in \S~\ref{obs}, the level of circular polarisation observed for the CO~$J=2-1$ emission is below the level that is considered significant in light of ALMA systematic circular polarisation calibration errors, we still compare the detected circular polarisation of $\sim0.4\%$ to the observed linear polarisation. As can be seen in Fig.~\ref{fig:zoom}, the circular polarisation is confined to the central part of the emission, with a size of approximately one interferometric beam. The linear polarisation vectors trace a smooth curved pattern through the region where there is a potential detection of circular polarisation. In the same region where the circular polarisation occurs, the direction of the continuum and CO linear polarisation is consistent (Fig.~\ref{fig:dustcomp}), which implies that there is no extra rotation of the polarisation vectors. We conclude that anisotropic resonant scattering does not affect our measurements.

\subsection{Magnetic field strength: Structure Function Analysis}
\label{sfa}

Similar to what is often done in studies of magnetic fields during star formation \citep[e.g.][]{Hildebrand09, Koch10, Houde16, DallOlio19} we can use the CO polarisation observations to provide an estimate of the magnetic field strength through a structure function analysis (SFA). The SFA analysis relates the dispersion of polarisation vectors at the smallest scales, to the turbulent motions in the magnetised gas to provide the ratio between the turbulent and mean large scale magnetic field strength. The SFA can thus be used to characterise turbulence. Under the assumption that the magnetic field is frozen into the gas, that the turbulent field arises from transverse Alfv{\'e}n waves, and that the turbulence is isotropic and incompressible, the ratio between the turbulent and mean large scale magnetic field strength determined using a SFA analysis can also be used to estimate the magnetic field strength.  

 In the following, we repeat the equations for the SFA from \citet{Koch10}. In the SFA, the dispersion of polarisation vectors is determined using the equation:
\begin{equation}
\langle\Delta\chi^2(l_k)\rangle^{1/2}\equiv \Biggl\{\frac{1}{N(l_k)}\sum^{N_k}_{l_k<r_{ij}<l_{k+1}} (\chi(r_i) - \chi(r_j))^2\Biggr\}^{1/2},    
\end{equation}
where $\chi(r_i)$ is the polarisation angle at position $r_i=(x_i,y_i)$, $r_{ij}=\sqrt{(x_i-x_j)^2+(y_i-y_j)^2}$, $l_k$ indicates the binning interval of the scale (taken in arcseconds), and $N(l_k)$ are the number of points averaged in each bin. It was shown in \citet{Hildebrand09} that, for a smooth magnetic field component $B_0$ and a small scale turbulent magnetic field component $B_t$, the structure function, on scales between the smallest turbulent scale and the characteristic length scale for variations in the large scale component, can be described with:
\begin{equation}
    \langle\Delta\chi^2(l_k)\rangle\simeq b^2 + m^2 l_k^2 + \sigma^2_M(l_k).
\end{equation}
Here $m$ is the slope of the linear dispersion term of the large scale field $B_0$ and $\sigma_M(l_k)$ propagates the observational uncertainties on the polarisation vectors in the binning. The term $b$ is then related to the ratio between the turbulent and large scale magnetic field strengths through:
\begin{equation}
    \frac{\langle B_t^2\rangle^{1/2}}{B_0} = \frac{b}{\sqrt{2-b^2}}.
    \label{eqsfa}
\end{equation}
Under the assumptions introduced previously, the ratio between the turbulent and large scale magnetic field component is equal to the ratio between the turbulent line width $\sigma_\nu$ and the Alfv{\'e}n velocity $\sigma_A=\frac{B_0}{\sqrt{4\pi\rho}}$, with $\rho$ the density of the gas. Using the turbulent velocity and average density in the CO emitting region, allows us to determine the magnetic field strength in the plane of the sky. This strength will be a lower limit in the case the magnetic field is not fully frozen into the gas.

The result of the SFA for two spectral channels, with the CO gas at $V_{\rm LSR}=57.3$~\kms and $62.4$~\kms around OH~17.7-2.0 are shown in
Fig.~\ref{fig:SFAOH}. As expected, the dispersion of the polarisation vectors increase from small to larger scale. For the channel close to the stellar velocity, the dispersion subsequently decreases steeply due to the aligned polarised emission at larger radius towards the North-East. The structure function is fit using Eq.~\ref{eqsfa} within $\lesssim2.3$\arcsec (three synthesised beams), which we take to correspond to a characteristic length scale for variations in the large scale magnetic field component. We note that for both velocity channels, the polarisation vector dispersion at the smallest scales available in our observations is $\sim20^\circ$. The analysis provides a ratio between the turbulent and large scale magnetic field component $\frac{\langle B_t^2\rangle^{1/2}}{B_0} = 0.10\pm0.01$, and $0.14\pm0.01$ for the two channels respectively. This means that the large scale magnetic field strength for both channels is:
\begin{equation}
    B_0=[2.3, 1.6]~\Biggl( \frac{\langle n_{H_2}\rangle}{10^5}\Biggr)^{1/2} \frac{v_t}{1.0}~{\rm mG}.
\end{equation}
 Here, $v_t$ is the turbulent velocity in the CO gas, taken to be $1.0$~\kms derived from CO radiative transfer modelling of AGB envelopes \citep[e.g.][]{Vlemmings21}. From similar models, the average H$_2$ number density $n_{H_2}$ in the CO region is assumed to be $10^5$~cm$^{-3}$. These assumptions, as well as the assumption that the magnetic field is frozen into the gas, dominate the uncertainty in the magnetic field strength. Similar results are obtained for the other channels for which a SFA could be performed. The polarised emission of the other molecular lines is limited to compact regions not much larger than our interferometric beam and hence a SFA was not possible.

 In \citet{Houde16} it is shown that a combination of the resolution of the observations and interferometric spatial filtering affect the results from the SFA. In \S\ref{obs} we showed that the maximum recoverable scale of our observations ($\sim8.1$\arcsec) is sufficiently large that spatial filtering does not affect our analysis. In order to determine the effect of the beamsize, we estimate the number of independent turbulent cells ($N$) probed by our observations using the formula from \citet{Houde16}:
\begin{equation}
    N = \frac{(\delta^2 + 2W_1^2)\Delta'}{\sqrt{2\pi}\delta^2}.
\end{equation}
Here $\delta$ is the correlation length of the turbulent field, $W_1$ is radius of the interferometric beam (we use $1/\sqrt{8\ln{2}}$ times an average fwhm beam-size of $0.72$\arcsec) and $\Delta'$ is the depth of the observed molecular layer along the line of sight. For a source at $D=2.94$~kpc, our beam radius $W_1\sim900$~au. We estimate the depth $\Delta'$ from the total size of the CO envelope ($\sim18000$~au) divided by the number of channels, yielding $\Delta'\sim1800$~au (although considering the sperically expanding CSE this value actually various across the envelope). Finally, we estimate the correlation length $\delta$ to be similar to the size of the molecular clumps that make up the envelope. \citet{Richards12} derived, from H$_2$O maser measurements, that these clumps start out with a size of the order of the stellar radius and expand for increasing distance to the star. Combined with CO observations and models \citep[e.g.][]{Olofsson96}, \citet{Richards12} suggest that the clump size $r_c$ appears to scale with distance from the star $r$ as $rc\propto r^0.8$. Applying these estimates to the CO~$J=2-1$ gas around OH~17.7-2.0 yields a size of $\sim850$~au. Hence, taking $\delta\sim850$~au, $W_1\sim900$~au and $\Delta'\sim1800$~au we find $N\sim2.7$. As the ratio between the turbulent and large scale magnetic field strengths scales approximately with $\sqrt{N}$, our magnetic field estimate would be overestimated by a factor of $\sim1.7$. Considering the various uncertainties in all the different assumptions, we conclude that the order of magnitude strength of the plane of the sky component of the magnetic field that permeates the CO emitting region is  $B_\perp\sim1$~mG.

 We can compare these results with the OH maser Zeeman measurements from B03. The Zeeman measurements of the paramagnetic OH molecule provide an estimate of the total magnetic field strength of $B=2.5-4.6$~mG. Since our measurement estimates the plane of the sky component of the magnetic field, the CO results are fully consistent with the OH Zeeman measurements.

\subsection{Comparison with OH linear polarisation results}
\label{OHcomp}

\begin{figure}[ht!]
\centering
\includegraphics[width=0.5\textwidth]{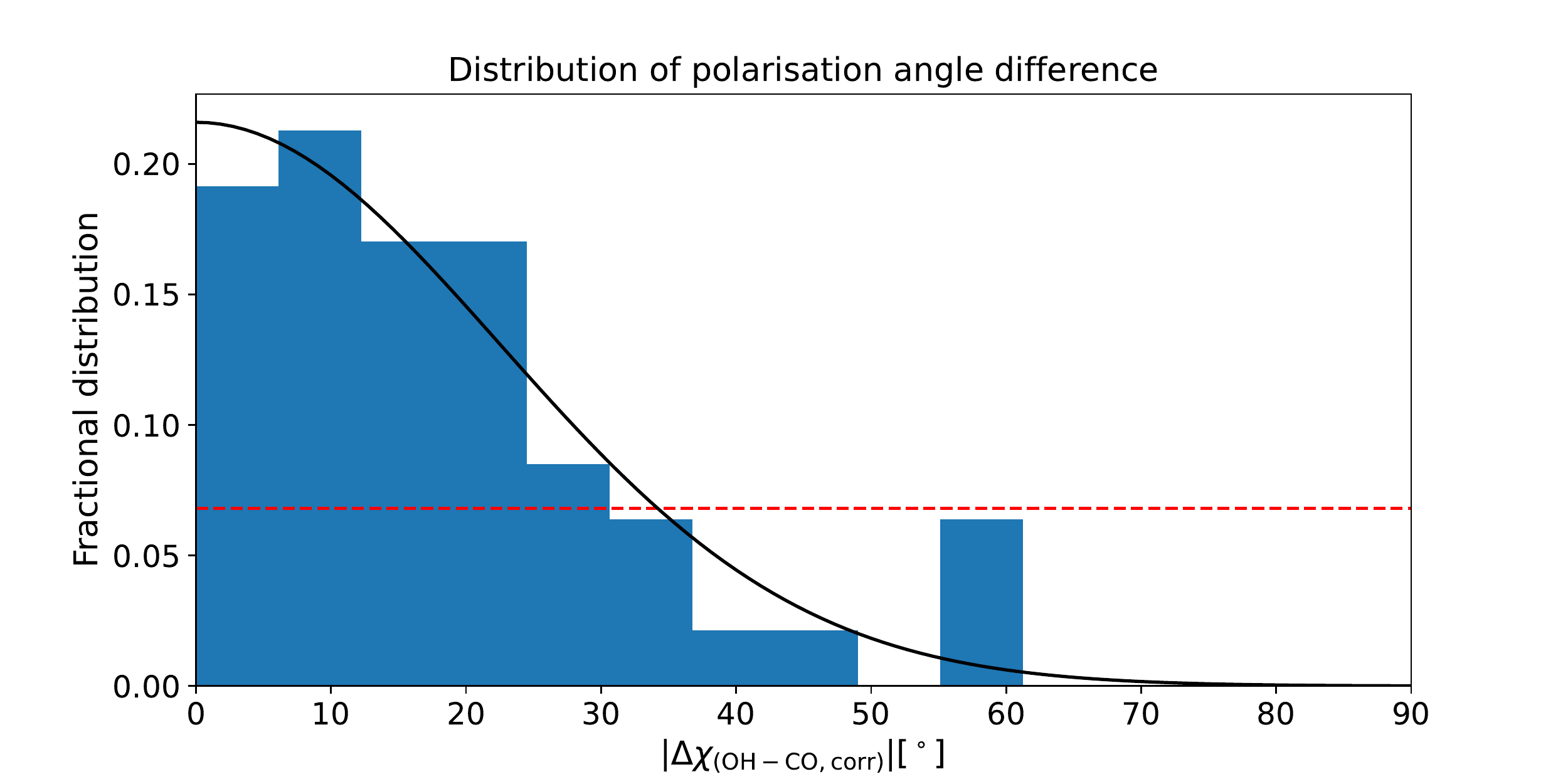}
\caption{The distribution of the absolute angle difference between the 1612~MHz OH masers (B03) and the CO polarisation direction. The mean difference, which can be affected by Faraday rotation, has been removed (see text). The solid black line indicates a normal distribution with $\sigma=22.5^\circ$, which corresponds to the best fit normal distribution. The dashed line represents a uniform distribution of angle differences.} 
\label{fig:angles}
\end{figure}

As previously indicated, a large scale magnetic field was mapped around OH~17.7-2.0 using OH maser observations (B03). This allows for a direct comparison between the magnetic field traced in individual 1612~MHz OH maser clumps and that in the more diffuse CO gas. The area covered by the OH masers is shown in a zoom-in of the three central velocity channels of our CO observations in Fig.~\ref{fig:zoom}. First, the maser positions where aligned with the CO observations by assuming that the centre of the maser shell fitted in B03 coincides with the peak of the continuum emission from our observations. Subsequently, each maser spot in table 4 of B03 was identified with the nearest $0.1$\arcsec pixel and nearest velocity channel in our observations. For those pixels that contained multiple maser spots, a weighted average OH maser polarisation angle was calculated. Subsequently, the mean angle was subtracted from both the OH polarisation angle ($\langle\chi_{\rm OH}\rangle=70.7^\circ$) and CO polarisation angle ($\langle\chi_{\rm CO}\rangle=81.6^\circ$) distributions. This was done because the OH maser polarisation angle is likely significantly affected by Faraday rotation (see below). The absolute angle difference $|\Delta\chi|=|(\chi_{\rm OH}-\langle\chi_{\rm OH}\rangle)-(\chi_{\rm CO}-\langle\chi_{\rm CO}\rangle)|$ is shown in Fig.~\ref{fig:angles}.

We find that the angle difference distribution can be best described by a normal distribution with a standard deviation $\sigma=22.5^\circ$. This is similar to the dispersion angle of the CO at the smallest scales as found in the SFA analysis in \S~\ref{sfa}. Hence, considering the similarity of the angle difference distribution with the dispersion in the angle expected to occur in both the OH masing and CO gas as a result of turbulence, we can confidently state that the magnetic field traced by the more diffuse CO gas and the OH maser clumps behaves similarly.

\subsection{Magnetic field morphology}
\label{morph}

The OH maser observations of B03 indicated that a large scale magnetic field is present around OH~17.7-2.0. The OH polarisation was attributed to emission from elliptically polarised $\sigma$-components that are the result of Zeeman splitting. This means that the polarisation vectors are perpendicular to the magnetic field direction. However, because foreground Faraday rotation is significant at OH maser frequencies, it was not possible to relate the polarisation direction with for example the outflow direction or inferred toroidal structure. Because the CO observations correspond to higher frequency emission ($\sim230$~GHz) compared to the OH maser observations at 1612~MHz, foreground Faraday rotation will be significantly less for our CO observations. We can estimate the Faraday rotation by using the best distance estimate ($2.94$~kpc, see \S~\ref{OH177}), a typical value for the interstellar magnetic field of $2~\mu$G \citep{Sun08} and an average electron number density $n_e=0.01066$~cm$^{-3}$ on the line of sight to OH~17.7-2.0 \citep{Yao17}. For the 1612~MHz OH masers, this implies a Faraday rotation of $\Delta\Phi=102^\circ$ while for CO the rotation would be negligible ($<0.005^{\circ}$).
Comparing this with the difference in mean angle between the OH and CO polarisation vectors found in \S~\ref{OHcomp} of $\sim -11^\circ$ this would imply an intrinsic angle difference of $\sim91^\circ$. This is remarkably close to perpendicular and strongly implies a relation between the polarisation measured on the masers and in the CO envelope. Considering the OH polarisation is perpendicular to the magnetic field, the CO polarisation is parallel to the magnetic field. Thus, the curvature in the CO polarisation from being perpendicular to the outflow to being nearly parallel to the outflow implies that we are tracing a dominant toroidal magnetic field component towards the South of the continuum peak and a helical or nearly poloidal field component towards the North part of the outflow. This is likely the result of the inclination of the outflow, which was estimated to be $\sim22^\circ$ with the blue-shifted outflow in the North and the red-shifted outflow in the South \citep{Gledhill11}. Towards the South we thus mostly probe gas in front of the outflow cavity, while in the North we probe part of the outflow cavity itself. Such a morphology is similar to that of magnetically driven outflow models such as a magnetic tower jet \citep[e.g.][]{Huarte12}, a configuration that was also inferred for the post-AGB object OH231.8+4.2 \citep{Sabin20}. As already noted in B03, the magnetic energy dominates the mechanical energy in the regions probed by the OH masers and considering the similar field strength estimated from the CO observations, the same holds true for the gas probed by the CO emission. 

\section{Conclusions}
\label{conc}

We have observed, using ALMA, molecular line polarisation in the CSE of the post-AGB star OH~17.7-2.0. The observations of the polarisation arising from the GK-effect, has allowed us to determine the magnetic field strength in the CO emitting region using a Structure Function Analysis. The strength of the magnetic field component in the plane of the sky was found to be $\sim1$~mG, which is consistent with previous Zeeman measurements using OH masers (B03). A comparison between the OH maser and CO linear polarisation vectors indicates that, although the OH maser polarisation direction is strongly affected by foreground Faraday rotation, the magnetic field in the OH masers and in the CO gas is likely the same. This confirms that the magnetic field properties, except for the absolute magnetic field direction on the sky, derived from OH masers are representative of the large scale circumstellar magnetic field. The magnetic field structure derived from the ALMA CO observations is similar to that expected for a magnetically driven outflow. As previously noted from the OH Zeeman observations, the magnetic energy also dominated the energy budget in the envelope. More detailed observations of the outflow launching region around OH~17.7-2.0 are needed to firmly identify the mechanism responsible for the structures seen around this post-AGB star. The observations presented here show that molecular line polarisation observations, of both maser and non-maser species, are invaluable to determine the role of magnetic fields during the late-stages of stellar evolution. 

\begin{acknowledgements}
  We thank the referee, Martin Houde, for comments that improved the paper.
  This paper makes use of the following ALMA data: ADS/JAO.ALMA\#2016.1.00251.S. ALMA is a partnership of ESO
  (representing its member states), NSF (USA) and NINS (Japan),
  together with NRC (Canada), NSC and ASIAA (Taiwan), and KASI
  (Republic of Korea), in cooperation with the Republic of Chile. The
  Joint ALMA Observatory is operated by ESO, AUI/NRAO and NAOJ. The project leading to this publication has received support from
ORP, that is funded by the European Union’s Horizon 2020 research
and innovation programme under grant agreement No 101004719 [ORP]. We
  also acknowledge support from the Nordic ALMA Regional Centre (ARC)
  node based at Onsala Space Observatory. The Nordic ARC node is
  funded through Swedish Research Council grant No 2017-00648.
\end{acknowledgements}

%
%

\bibliographystyle{aa}

\begin{thebibliography}{30}
\expandafter\ifx\csname natexlab\endcsname\relax\def\natexlab#1{#1}\fi

\bibitem[Bains et al.(2003)]{Bains03} Bains, I., Gledhill, T.~M., Yates, J.~A., et al.\ 2003, \mnras, 338, 287 (B03)

\bibitem[Beuther et al.(2010)]{Beuther10} Beuther, H., Vlemmings, W.~H.~T., Rao, R., et al.\ 2010, \apjl, 724, L113

\bibitem[Blackman(2022)]{Blackman22} Blackman, E.~G.\ 2022, arXiv:2202.07246

\bibitem[Boffin \& Jones(2019)]{Boffin19} Boffin, H.~M.~J. \& Jones, D.\ 2019, The Importance of Binaries in the Formation and Evolution of Planetary Nebulae, SpringerBriefs in Astronomy. ISBN 978-3-030-25058-4. The Author(s), under exclusive license to Springer Nature Switzerland AG, 2019

\bibitem[Bowers(1978)]{Bowers78} Bowers, P.~F.\ 1978, \aaps, 31, 127

\bibitem[Bowers et al.(1983)]{Bowers83} Bowers, P.~F., Johnston, K.~J., \& Spencer, J.~H.\ 1983, \apj, 274, 733

\bibitem[Cort{\'e}s et al.(2005)]{Cortes05} Cort{\'e}s, P.~C., Crutcher, R.~M., \& Watson, W.~D.\ 2005, \apj, 628, 780

\bibitem[Cort{\'e}s et al.(2021)]{Cortes21} Cort{\'e}s, P.~C., Sanhueza, P., Houde, M., et al.\ 2021, \apj, 923, 204

\bibitem[Dall'Olio et al.(2019)]{DallOlio19} Dall'Olio, D., Vlemmings, W.~H.~T., Persson, M.~V., et al.\ 2019, \aap, 626, A36

\bibitem[Engels(2002)]{Engels02} Engels, D.\ 2002, \aap, 388, 252

\bibitem[Gaia Collaboration et al.(2021)]{Gaia21} Gaia Collaboration, Brown, A.~G.~A., Vallenari, A., et al.\ 2021, \aap, 649, A1

\bibitem[Girart et al.(2012)]{Girart12} Girart, J.~M., Patel, N., Vlemmings, W.~H.~T., et al.\ 2012, \apjl, 751, L20

\bibitem[Gledhill et al.(2011)]{Gledhill11} Gledhill, T.~M., Forde, K.~P., Lowe, K.~T.~E., et al.\ 2011, \mnras, 411, 1453

\bibitem[Goldreich \& Kylafis(1982)]{GK82} Goldreich, P. \& Kylafis, N.~D.\ 1982, \apj, 253, 606

\bibitem[Gonidakis et al.(2014)]{Gonidakis14} Gonidakis, I., Chapman, J.~M., Deacon, R.~M., et al.\ 2014, \mnras, 443, 3819

\bibitem[Herman \& Habing(1985)]{HH85} Herman, J. \& Habing, H.~J.\ 1985, \aaps, 59, 523

\bibitem[Herpin et al.(2006)]{Herpin06} Herpin, F., Baudry, A., Thum, C., et al.\ 2006, \aap, 450, 667

\bibitem[Heske et al.(1990)]{Heske90} Heske, A., Forveille, T., Omont, A., et al.\ 1990, \aap, 239, 173

\bibitem[Hildebrand et al.(2009)]{Hildebrand09} Hildebrand, R.~H., Kirby, L., Dotson, J.~L., et al.\ 2009, \apj, 696, 567

\bibitem[Houde et al.(2013)]{Houde13} Houde, M., Hezareh, T., Jones, S., et al.\ 2013, \apj, 764, 24

\bibitem[Houde et al.(2016)]{Houde16} Houde, M., Hull, C.~L.~H., Plambeck, R.~L., et al.\ 2016, \apj, 820, 38

\bibitem[Houde et al.(2022)]{Houde22} Houde, M., Lankhaar, B., Rajabi, F., et al.\ 2022, \mnras, 511, 295

\bibitem[Huang et al.(2020)]{Huang20} Huang, K.-Y., Kemball, A.~J., Vlemmings, W.~H.~T., et al.\ 2020, \apj, 899, 152

\bibitem[Huarte-Espinosa et al.(2012)]{Huarte12} Huarte-Espinosa, M., Frank, A., Blackman, E.~G., et al.\ 2012, \apj, 757, 66

\bibitem[Koch et al.(2010)]{Koch10} Koch, P.~M., Tang, Y.-W., \& Ho, P.~T.~P.\ 2010, \apj, 721, 815

\bibitem[Khouri et al.(2021)]{Khouri22} Khouri, T., Vlemmings, W.~H.~T., Tafoya, D., et al.\ 2021, Nature Astronomy, 6, 275

\bibitem[Lagadec et al.(2011)]{Lagadec11} Lagadec, E., Verhoelst, T., M{\'e}karnia, D., et al.\ 2011, \mnras, 417, 32

\bibitem[Lankhaar \& Vlemmings(2020)]{Lankhaar20} Lankhaar, B. \& Vlemmings, W.\ 2020, \aap, 636, A14

\bibitem[Le Bertre et al.(1984)]{LeBertre84} Le Bertre, T., Epchtein, N., \& Nguyen-Q-Rieu\ 1984, \aap, 138, 353

\bibitem[Leal-Ferreira et al.(2013)]{LealFerreira13} Leal-Ferreira, M.~L., Vlemmings, W.~H.~T., Kemball, A., et al.\ 2013, \aap, 554, A134
  
\bibitem[L{\`e}bre et al.(2014)]{Lebre14} L{\`e}bre, A., Auri{\`e}re, M., Fabas, N., et al.\ 2014, \aap, 561, A85

\bibitem[Li \& Henning(2011)]{Li11} Li, H.-B. \& Henning, T.\ 2011, \nat, 479, 499

\bibitem[{{McMullin} {et~al.}(2007){McMullin}, {Waters}, {Schiebel}, {Young},
  \& {Golap}}]{McMullin07}
{McMullin}, J.~P., {Waters}, B., {Schiebel}, D., {Young}, W., \& {Golap}, K.
  2007, in Astronomical Society of the Pacific Conference Series, Vol. 376,
  Astronomical Data Analysis Software and Systems XVI, ed. R.~A. {Shaw},
  F.~{Hill}, \& D.~J. {Bell}, 127

\bibitem[Murakawa et al.(2013)]{Murakawa13} Murakawa, K., Izumiura, H., Oudmaijer, R.~D., et al.\ 2013, \mnras, 430, 3112

\bibitem[Nagai et al.(2016)]{Nagai16} Nagai, H., Nakanishi, K., Paladino, R., et al.\ 2016, \apj, 824, 132

\bibitem[Nordhaus \& Blackman(2006)]{Nordhaus06} Nordhaus, J. \& Blackman, E.~G.\ 2006, \mnras, 370, 2004

\bibitem[Olofsson et al.(1996)]{Olofsson96} Olofsson, H., Bergman, P., Eriksson, K., et al.\ 1996, \aap, 311, 587

\bibitem[Ondratschek et al.(2022)]{O22} Ondratschek, P.~A., R{\"o}pke, F.~K., Schneider, F.~R.~N., et al.\ 2022, \aap, 660, L8.

\bibitem[Perez-Sanchez et al.(2013)]{PerezSanchez13} Perez-Sanchez, A.~F., Vlemmings, W.~H.~T., Tafoya, D., et al.\ 2013, \mnras, 436, L79

\bibitem[Reid et al.(2014)]{Reid14} Reid, M.~J., Menten, K.~M., Brunthaler, A., et al.\ 2014, \apj, 783, 130

\bibitem[Richards et al.(2012)]{Richards12} Richards, A.~M.~S., Etoka, S., Gray, M.~D., et al.\ 2012, \aap, 546, A16

\bibitem[Sabin et al.(2015)]{Sabin15} Sabin, L., Wade, G.~A., \& L{\`e}bre, A.\ 2015, \mnras, 446, 1988

\bibitem[Sabin et al.(2020)]{Sabin20} Sabin, L., Sahai, R., Vlemmings, W.~H.~T., et al.\ 2020, \mnras, 495, 4297

\bibitem[S{\'a}nchez Contreras et al.(2007)]{SC07} S{\'a}nchez Contreras, C., Le Mignant, D., Sahai, R., et al.\ 2007, \apj, 656, 1150

\bibitem[S{\'a}nchez Contreras \& Sahai(2012)]{SCS12} S{\'a}nchez Contreras, C. \& Sahai, R.\ 2012, \apjs, 203, 16

\bibitem[Soker(2002)]{Soker02} Soker, N.\ 2002, \mnras, 336, 826

\bibitem[Stephens et al.(2020)]{Stephens20} Stephens, I.~W., Fern{\'a}ndez-L{\'o}pez, M., Li, Z.-Y., et al.\ 2020, \apj, 901, 71

\bibitem[Sun et al.(2008)]{Sun08} Sun, X.~H., Reich, W., Waelkens, A., et al.\ 2008, \aap, 477, 573

\bibitem[Teague et al.(2021)]{Teague21} Teague, R., Hull, C.~L.~H., Guilloteau, S., et al.\ 2021, \apj, 922, 139

\bibitem[Vickers et al.(2015)]{Vickers15} Vickers, S.~B., Frew, D.~J., Parker, Q.~A., et al.\ 2015, \mnras, 447, 1673

\bibitem[Vlemmings et al.(2002)]{Vlemmings02} Vlemmings, W.~H.~T., Diamond, P.~J., \& van Langevelde, H.~J.\ 2002, \aap, 394, 589

\bibitem[Vlemmings et al.(2005)]{Vlemmings05} Vlemmings, W.~H.~T., van Langevelde, H.~J., \& Diamond, P.~J.\ 2005, \aap, 434, 1029

\bibitem[Vlemmings et al.(2006)]{Vlemmings06} Vlemmings, W.~H.~T., Diamond, P.~J., \& Imai, H.\ 2006, \nat, 440, 58

\bibitem[Vlemmings et al.(2012)]{Vlemmings12} Vlemmings, W.~H.~T., Ramstedt, S., Rao, R., et al.\ 2012, \aap, 540, L3

\bibitem[Vlemmings(2019)]{Vlemmings19} Vlemmings, W.\ 2019, IAU Symposium, 343, 19

\bibitem[Vlemmings et al.(2017)]{Vlemmings17} Vlemmings, W.~H.~T., Khouri, T., Mart{\'\i}-Vidal, I., et al.\ 2017, \aap, 603, A92

\bibitem[Vlemmings et al.(2021)]{Vlemmings21} Vlemmings, W.~H.~T., Khouri, T., \& Tafoya, D.\ 2021, \aap, 654, A18

\bibitem[Wolak et al.(2013)]{Wolak13} Wolak, P., Szymczak, M., Bartkiewicz, A., et al.\ 2013, The Astronomer's Telegram, 5211

\bibitem[Yao et al.(2017)]{Yao17} Yao, J.~M., Manchester, R.~N., \& Wang, N.\ 2017, \apj, 835, 29

\end{thebibliography}

\begin{appendix}
\section{Polarisation maps for three further molecules}

\begin{figure*}[ht!]
\centering
\includegraphics[width=0.95\textwidth]{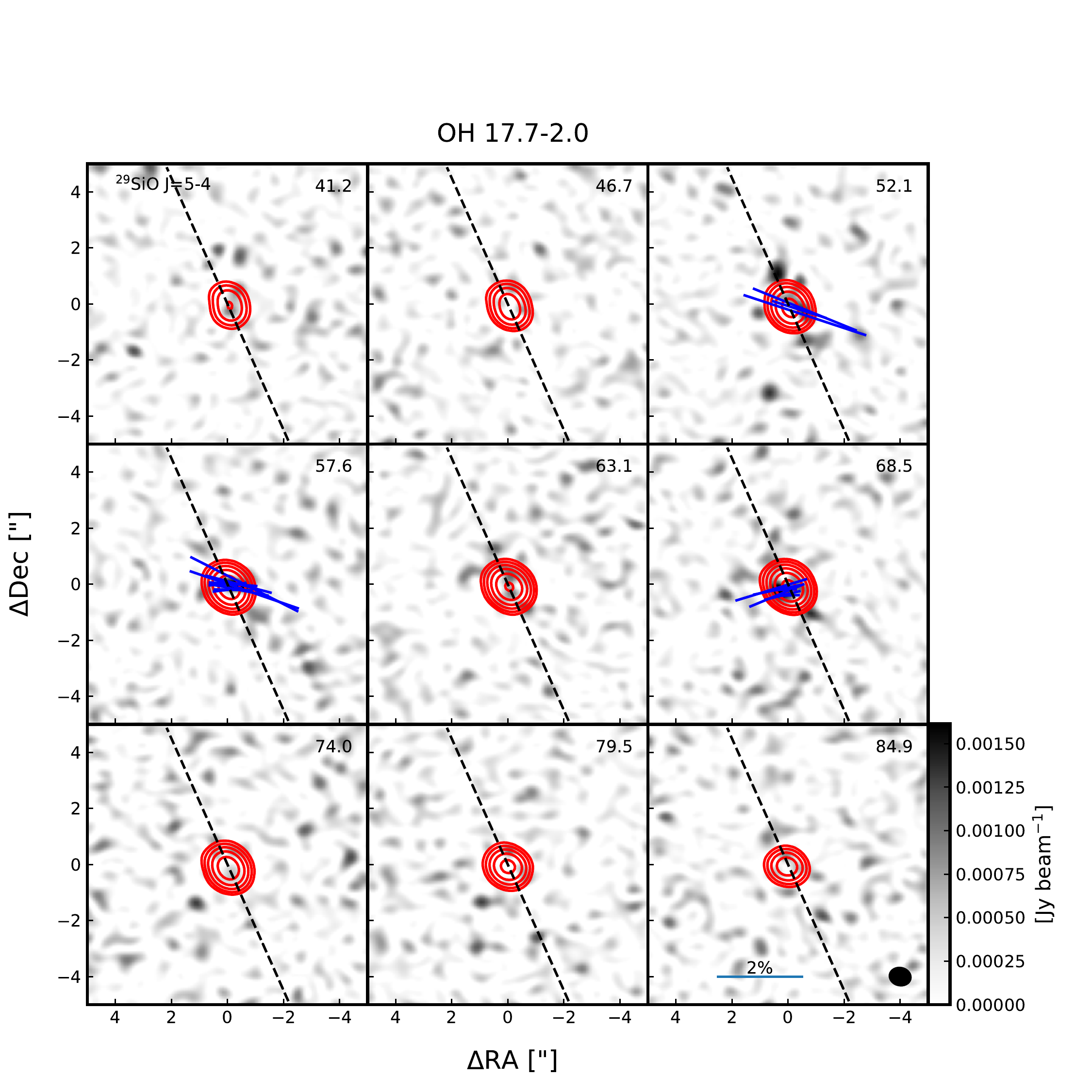}
\caption{Same as Fig.~\ref{fig:CO} for the $^{29}$SiO $v=0, J=5-4$ emission around OH~17.7-2.0. The peak emission is $I_{\rm 29SiO, peak}$=0.31~Jy~beam$^{-1}$. The maximum polarisation fraction $P_{\rm l, max}= 4.7\%$.}
\label{fig:29SiO}
\end{figure*}

\begin{figure*}[ht!]
\centering
\includegraphics[width=0.95\textwidth]{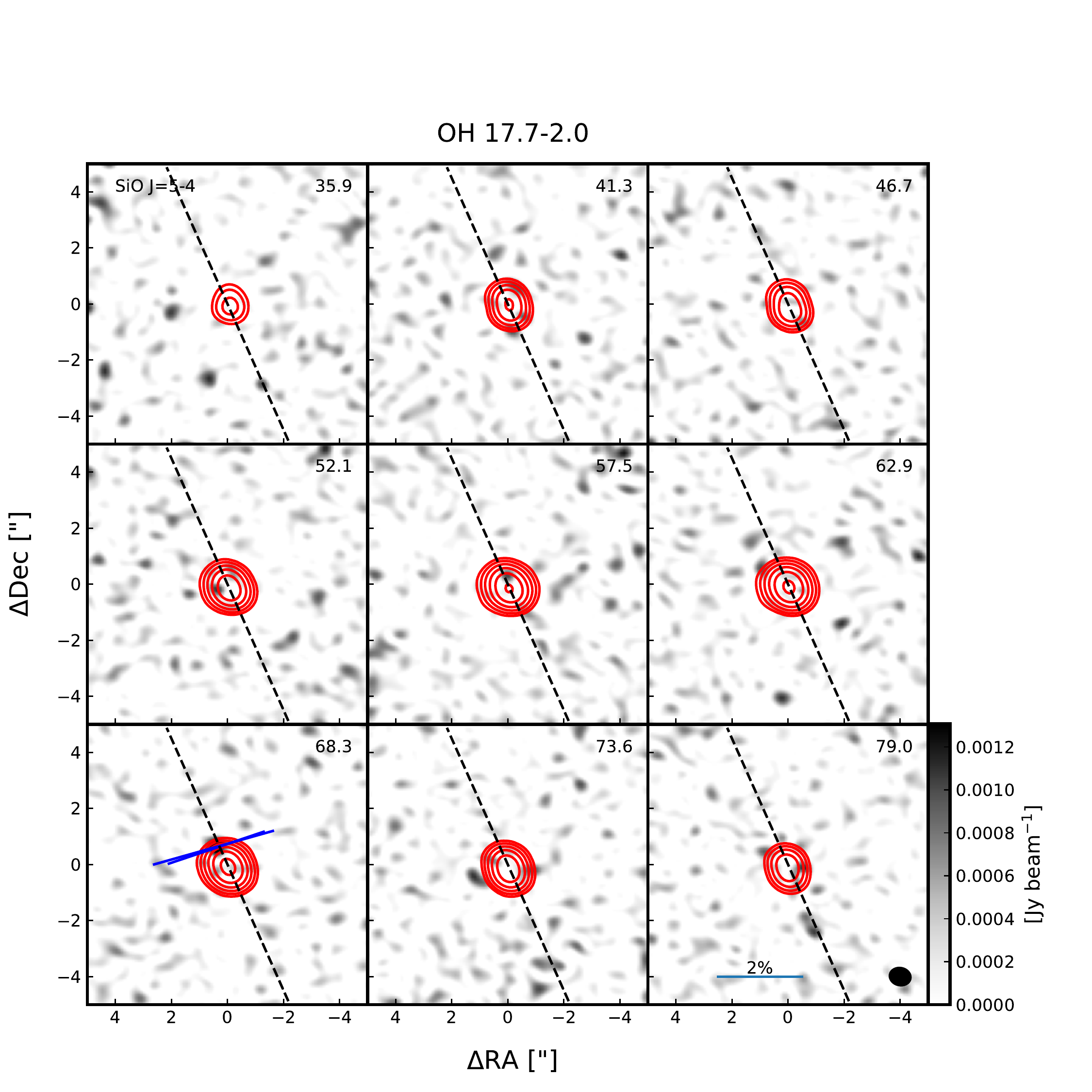}
\caption{Same as Fig.~\ref{fig:CO} for the SiO $v=0, J=5-4$ emission around OH~17.7-2.0. The peak emission is $I_{\rm SiO, peak}$=0.49~Jy~beam$^{-1}$ and the maximum polarisation fraction $P_{\rm l, max}= 2.9\%$.}
\label{fig:SiO}
\end{figure*}

\begin{figure*}[ht!]
\centering
\includegraphics[width=0.95\textwidth]{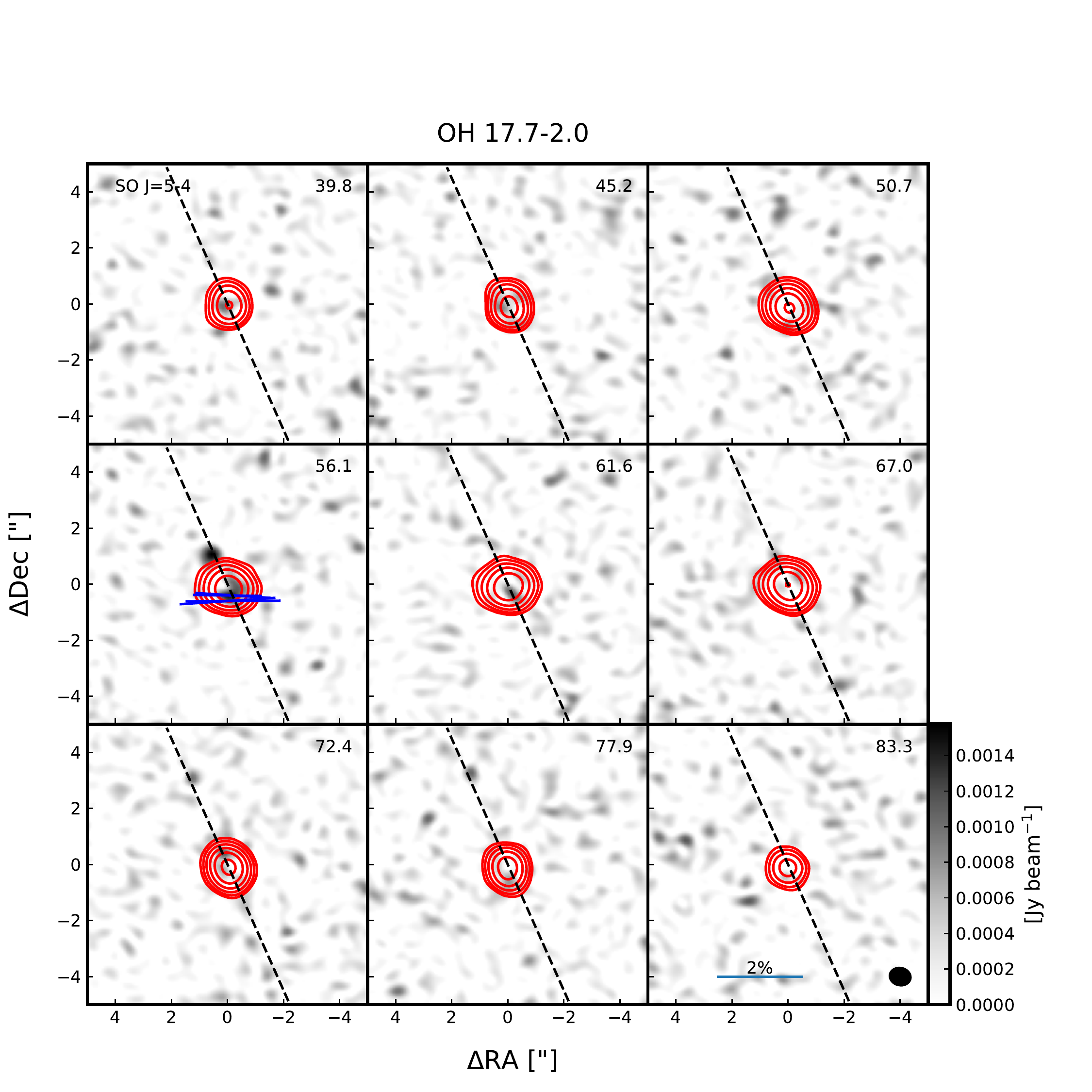}
\caption{Same as Fig.~\ref{fig:CO} for the SO $J=5-4$ emission around OH~17.7-2.0. The peak emission is $I_{\rm SO, peak}$=0.12~Jy~beam$^{-1}$ and the maximum polarisation fraction $P_{\rm l, max}= 2.2\%$.}
\label{fig:SO54}
\end{figure*}

\end{appendix}

\end{document}